# Stable Routing for achieving Quality of Service in wireless Sensor Networks

Deepali Virmani
B5/107 Mayur Apartment, Sector 9
Rohini, Delhi, India.
deepalivirmani@gmail.com

Satbir Jain
NSIT, Secto 3, Dwarka
Delhi,India
jain_satbir@yahoo.com

**Abstract.** Networking in Wireless Sensor networks is a challenging task due to the lack of resources in the network as well as the frequent changes in network topology. Although lots of research has been done on supporting QoS in the Internet and other networks, but they are not suitable for wireless sensor networks and still QoS support for such networks remains an open problem. In this paper, a new scheme has been proposed for achieving QoS in terms of packet delivery, multiple connections, better power management and stable routes in case of failure. It offers quick adaptation to distributed processing, dynamic linking, low processing overhead and loop freedom at all times. The proposed scheme has been incorporated using QDPRA protocol and by extensive simulation the performance has been studied, and it is clearly shown that the proposed scheme performs very well for different network scenarios.

## 1 Introduction and Motivation

A wireless sensor network is a collection of sensor nodes equipped with interfaces and networking capability [1]. Such devices can communicate with another node within their radio range or one that is outside their range by multi hop techniques. Wireless sensor network is adaptive in nature and is self organizing. In this wireless topology may change rapidly and unpredictably. The main characteristic of WSN strictly depends upon both wireless link nature and node mobility features. Basically this includes dynamic topology, bandwidth, energy constraints, security limitations and lack of infrastructure. WSN's are viewed as suitable systems which can support some specific applications as virtual classrooms, military communications, emergency search and rescue operations, data acquisition in hostile environments, communications set up in Exhibitions, conferences and meetings, in battle field among soldiers to coordinate defense or attack, at airport terminals for workers to share files etc. Several routing protocols for WSN's have been proposed in the literature. In most of the routing protocols, major emphasis has been on finding shortest routes. In this paper a new scheme the stable routing by using virtual nodes for self stabilization with power factor (SRVNP) has been suggested which would allow sensor nodes to maintain routes to destinations with more stable route selection. This scheme responds to link breakages and changes in network topology in a timely manner. Routing in wireless sensor networks experiences more link failures than in other networks. Hence, a routing protocol that supports QoS for Wireless sensor networks requires considering the reasons for link failure to improve its performance. Link failure stems from node mobility and lack of the network resources. Therefore it is essential to capture the aforesaid characteristics to identify the quality of links. Furthermore, the routing protocols must be adaptive to cope with the time-varying low-capacity resources. For instance, it is possible that a route that was earlier found to meet certain QoS requirements no longer does so due to the dynamic nature of the topology. In such a case, it is important that the network intelligently adapts the session to its new and changed conditions. Quality of service [2] means providing a set of service requirements to the flows while routing them through the network. A new scheme has been suggested which combines two basic features to achieve QoS; these are stable routing and concept of battery power as the battery is main concern with WSN's. The scheme uses virtual ndes for stable routes and uses power factor to determine active nodes to participate in routing. The rest of the paper is organized as follows: Section 2 analyzes new proposed scheme (SRVNP), Section 3 describes the simulation environment and results and Section 4 summarizes the study and the status of the work.

## 2. Proposed Scheme: SRVNP

The proposed scheme takes care of on demand routing along with a new concept of virtual nodes with power factor. Many protocols have been discussed using concept of power in many existing schemes [3-11]. In all the schemes discussed under concept of power routing, no concern has been taken for stable routing or better packet delivery. All emphasis is on concept of battery power or energy requirement for routing process. In this paper two different concepts have been joined together to make an efficient protocol. Major concentration is on the routing problem. In the proposed scheme, the virtual nodes help in reconstruction phase in fast selection of new routes. Selection of virtual nodes is made upon availability of nodes and battery status. Each route table has an entry for number of virtual nodes attached to it and their battery status. The protocol is divided into three phases. Route Request (RReq), Route Repair (RRpr) and Error Phase (Err).

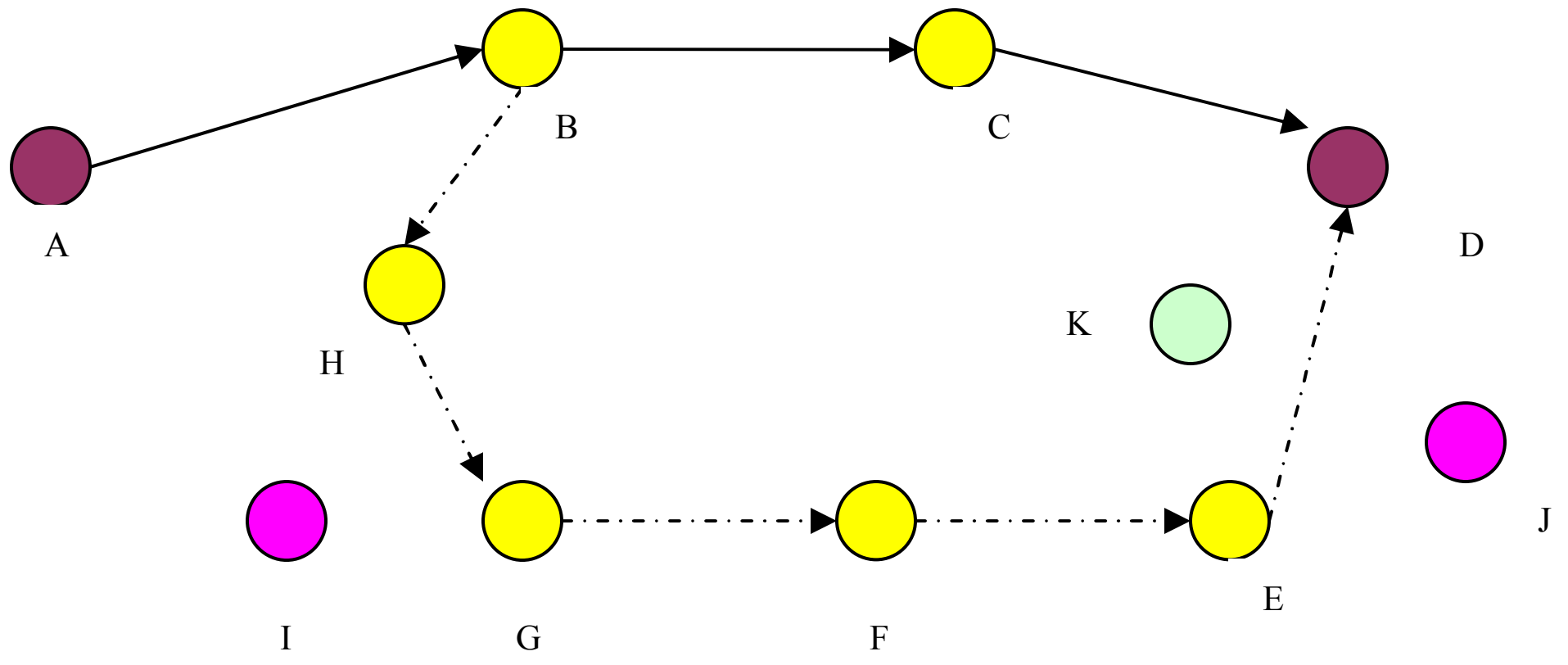

**Fig.1.** An example of routing

The proposed scheme is explained with the help of an example shown in Figure 1. Assume that the node A is the source while destination is the node D. Note that the route discovered using new scheme routing protocol may not necessarily be the shortest route between a source destination pair. If the node C is having power status in critical or danger zone, then though the shortest path is A-B-C-D but the more stable path A-B-H-G-F-E-D in terms of active power status is chosen. This may lead to slight delay but improves overall efficiency of the protocol by sending more packets without link break than the state when some node is unable to process route due to inadequate battery power. The process also helps when some intermediate node moves out of the range and link break occurs, in that case virtual nodes take care of the process and the route is established again without much overhead. In Figure 1 if the node G moves out, the new established route will be A-B-H-I-F-E-D. Here the node I is acting as virtual node(VN) for the node H and the node G. Similarly the node J can be VN for the nodes D,E,K. Virtual node(VN) has been selected at one hop distance from the said node.

**2.1 Route Construction (RReq) Phase**

This scheme can be incorporated with reactive routing protocols that build routes on demand via a query and reply procedure. The scheme does not require any modification to the QDPRA's[12] RReq (route request) propagation process. In this scheme when a source needs to initiate a data session to a destination but does not have any route information, it searches a route by flooding a ROUTE REQUEST (RReq) packet. Each RReq packet has a unique identifier so that nodes can detect and drop duplicate packets. An Intermediate node with an active route (in terms of power and Vitual Nodes), upon receiving a no duplicate RReq, records the previous hop and the source node information in its route table i.e. backward learning. It then broadcasts the packet or sends back a ROUTE REPLY (RRep) packet to the source if it has an active route to the destination. The destination node sends a RRep via the selected route when it receives the first RReq or subsequent RReq's that traversed a better active route. Nodes monitor the link status of next hops in active routes. When a link break in an active route is detected, an Err message is used to notify that the loss of link has occurred to its one hop neighbor. Here Err message indicates those destinations which are no longer reachable. Taking advantage of the broadcast nature of wireless communications, a node promiscuously overhears packets that are transmitted by their neighboring nodes. When a node that is not part of the route overhears a RRpr packet not directed to itself transmit by a neighbor (on the primary route), it records that neighbor as the next hop to the destination in its alternate route table. From these packets, a node obtains alternate path information and makes entries of these virtual nodes (VN) in its route table. If route breaks occurs it just starts route construction phase from that node. The protocol updates list of VNs and their power status periodically in the route table.

### 2.2 Route Error & Maintenance

In this scheme data transmits continuously through the primary route unless there is a route disconnection. When a node detects a link break, it performs a one hop data broadcast to its immediate neighbors. The node specifies in the data header that the link is disconnected and thus the packet is candidate for alternate routing. Upon receiving this packet route maintenance phase starts by selecting alternate path and checking power status.

### 2.3 Local Route Repair (Err Phase)

When a link break in an active route occurs, the node upstream of that break may choose to repair the link locally if the destination was no farther and there exists VNs that are active. The Time to live (TTL) of the RReq should initially be set to the following value:

$$TTL = \max(Min_Rpr_TTL + VN, 0.5 * \#hops) + power\,status \quad (1)$$

Where Min_Rpr_TTL is the last known hop count to destination, #hops is the number of hops to the sender of the currently undeliverable packets. VN is the virtual nodes attached to the said node and the power status is power state of the node at that time. As 0.5* #hops is always less than Min_Rpr_TTL + VN , so the whole process becomes invisible to the originating node.

This factor is transmitted to all nodes to select best available path with maximum power.

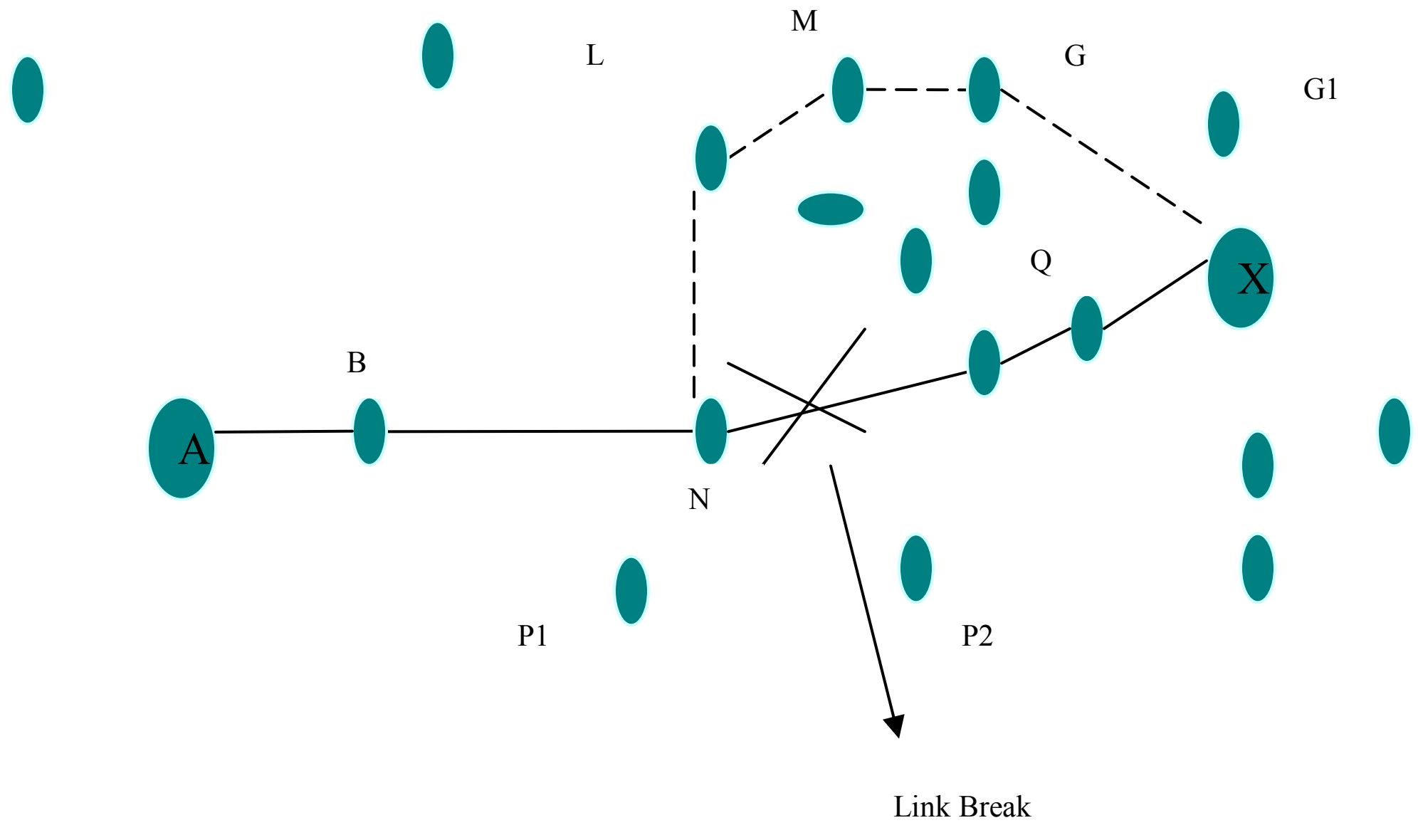

**Fig. 2.** Local Repair

Figure 2 gives an idea of working of local route repair. Initial path from source node A to destination node X is shown via solid lines. When link breaks at node N, route repair starts, node N starts searching for new paths, buffering packets from A-B in its buffer. Node N invokes Route Request phase for X
.

**Table 1.** Active Time Estimation

| Node | VN | Min_TTL | #hops*0.5 | Power Status | Total |
|---|---|---|---|---|---|
| L | 3 | 3 | 1/2 | 9 | 15 |
| M | 4 | 2 | 2/2 | 8.5 | 14.5 |
| G | 3 | 1 | 3/2 | 8 | 12 |
| P | 3 | 1 | 2/2 | 4 | 8 |
| Q | 3 | 1 | 3/2 | 3 | 7 |
| P1 | 1 | 4 | 1/2 | 7 | 12.5 |
| P2 | 2 | 3 | 2/2 | 7 | 12 |
| G1 | 2 | 1 | 4/2 | 7.5 | 10.5 |
| L1 | 1 | 3 | 2/2 | 8.0 | 12 |

Now backbone nodes are selected and proper selection of nodes is done based on power factor. Path selected becomes [N-L-M-K-X], instead of [N-L-P-X], since the node P is not in active state. Even though the route may become longer, but the selected route path is far more stable and delivers more packets. Stability of route depends upon two major aspects as: Life time and Power status. The concept has been explained in Table 1.

When selection has to be made between nodes P1 and L at the start of repair phase, selection of node L has the advantage over node P1. Similarly in the selection between nodes K and K1, node K has higher weight. If any VN has not been on active scale, it is rejected and a new node is searched. In addition to power factor, efforts are made to keep the path shortest. This local repair attempts are often invisible to the originating node. During local repair data packets will be buffered at local originator. If, at the end of the discovery period, the repairing node has not received a reply message RRpr it proceeds in by transmitting a route error Err to the originating node. On the other hand, if the node receives one or more route reply RRep's during the discovery period, it first compares the hop count of the new route with the value in the hop count field of the invalid route table entry for that destination. Repairing the link locally is likely to increase the number of data packets that are able to be delivered to the destinations, since data packets will not be dropped as the ERR travels to the originating node. Sending a ERR to the originating node after locally repairing the link break may allow the originator to find a fresh route to the destination that is better, based on current node positions. However, it does not require the originating node to rebuild the route, as the originator may be done, or nearly done, with the data session. In AODV, a route is timed out when it is not used and updated for certain duration of time. The proposed scheme uses the same technique for timing out alternate routes.

## 3. Simulation and Results

Simulation study has been carried out to study the Performance study of existing different protocols. Simulation Environment used is J-Sim(Java simulator) to carry out the process. Simulation results have been compared with QDPRA, AODV, DSR and TORA. Simulation study has been performed for packet delivery ratio, Throughput and End to End delay evaluations.

**Packet Delivery Ratio:** The fraction of successfully received packets, which survive while finding their destination. This performance measure also determines the completeness and correctness of the routing protocol. If F is total flows , i is node , PR is packets received from i , PT is transmitted from i, then $D_R$ (Delivery Ratio) can be determined by

$$D_R = \frac{1}{F}\sum_{i=1}^{F}\frac{P_{Ri}}{T_i} \qquad (2)$$

**End-to-End Delay:** Average end-to-end delay is the delay experienced by the successfully delivered packets in reaching their destinations. This is a good metric for comparing protocols and denotes how efficient the underlying routing algorithm is, because delay primarily depends on optimality of path chosen.

Let A is the total packets transmitted, $r_i$ is the number of packets received successfully by node i and $A_i$ is the total packets transmitted by node i ,then *AD*( E to E Delay ) can be determined by

$$AD = \frac{1}{A}\sum_{i=1}^{A} r_i - A_i \tag{3}$$

**Throughput:** It is defined as rate of successfully transmitted data per second in the network during the simulation. Throughput is calculated such that, it is the sum of successfully delivered payload sizes of data packets within the period, which starts when a source opens a communication port to a remote destination port, and which ends when the simulation stops. Average throughput can be calculated by dividing total number of bytes received by the total end-to-end delay.

### 3.1 Results

**Packet Delivery Ratio:** In simulation study 100 nodes were taken in a random scenario of size 1000 × 1000. Two parameters have been takes as Pause time and speed. The study has been conducted at different pause times. Pause time of 0 means maximum mobility and 500 is minimum mobility. The sources connected are 25-34 using TCP connection. Figure 3 represents the results. DSR is not delivering more than 84% in denser mediums. It is unable to perform better in higher congestion zones. AODV outperforms DSR in congested medium. DSR drops significant packets at high speed; the reason can be use of source routing and aggressive use of cache. With higher loads the extent of cache is too large to benefit performance. Often stale routes are chosen and all this leads to more packet falls. AODV is delivering more packets to DSR in most of the cases and has an edge over it. QDPRA performance is better than AODV and DSR .New scheme (SRVNP) is overall best for 100 nodes. It starts with 86% and with increasing pause time gets stable and delivers more than 95% packets. Figure 4 shows the simulation results with speed as a function. AODV and DSR have performed better at all speeds. DSR cache performance has suffered a bit at higher speed in denser medium. The reason is that keeping cache for such a large network demand more storage and in turn slows packet delivery rate. DSR is able to deliver between 90% to 94% packets all the time. AODV improves in denser mediums as it is able to support more packets. It overpowers DSR at high speed of 15 t20 meters per second and trend is true even at higher speeds. Delivery rate of QDPRA is between 92% to 98%. Proposed scheme has been the best in dense mediums, showing almost same performance at all speeds. In case of new scheme delivery ratio was nearing 98% even at higher speeds 10 to 20 and more. Proposed scheme outperforms all other schemes. This proves New scheme performs better in denser medium, as more virtual nodes are available for route selection and also more nodes are available with better power status.

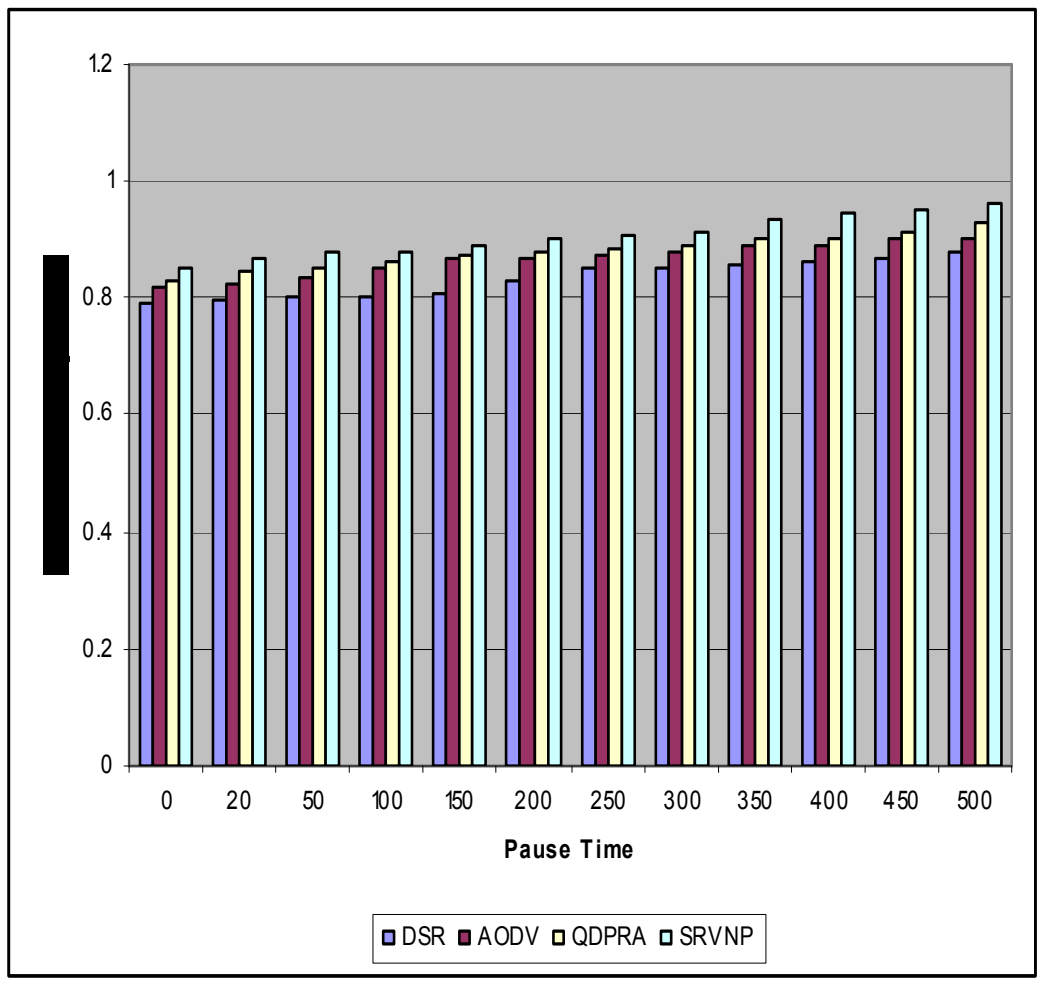

**Fig.3.** Packet Delivery ratio at different pause time

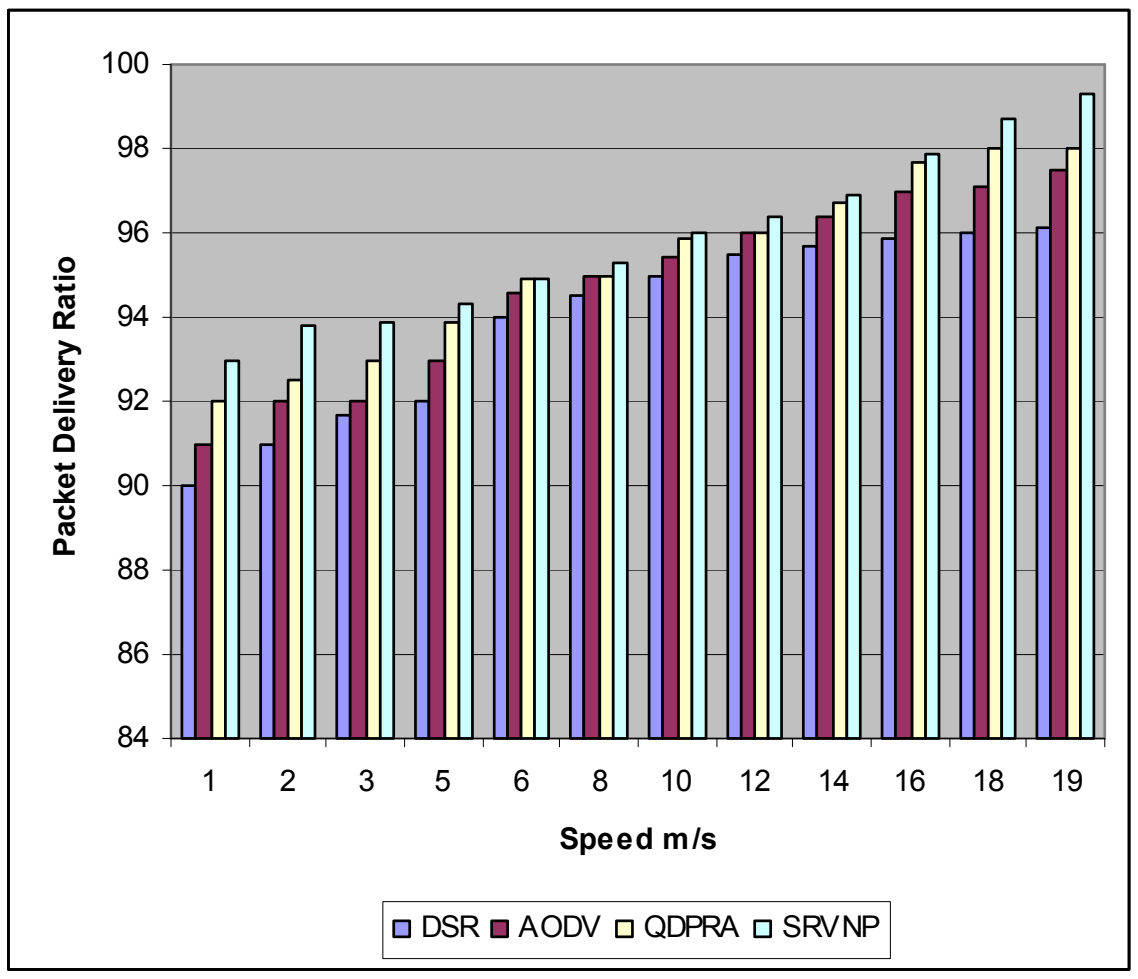

**Fig. 4.** Packet Delivery ratio at different speed

**End to End delay** has been explained in Figure 5. Here it is clear that DSR has more delays than AODV. The protocol proposed has higher delays. While DSR uses source routing, it gains so much information by the source that it will learn routes to many destinations than an distance vector protocol like AODV or New. This means while DSR already has a route for a certain destination, New would have to send a specific request for that destination. The packets would in the meanwhile stay in the buffer until a valid route is found. This takes some time and will therefore increases the average delay. Delay for QDPRA is still more than DSR. The delay for SRVNP is more and the reason is that it spends more time in calculation of stable route. New does deliver even those packets, which may have been dropped in AODV as it has better recovery mechanism and local repair system for faster recovery. All this process increases delay but not at the cost of efficiency.

**Throughput** in bytes per second has been calculated and speed has been varied from 0 to 3.5 meter per second. Figure 6 shows the graphical representation. DSR, AODV and QDPRA and SRVNP have an increase in throughput. The throughput increase can be further explained by TCP behavior, such that, when ACK is not received back, TCP source retransmits data. With increased mobility, source may distribute several identical packets to different intermediate nodes, because of route changes At 1.5 m/s speed, AODV protocol also shows a decreasing behavior with the increased network speed. But SRVNP shows increase in throughput even if speed is increased.

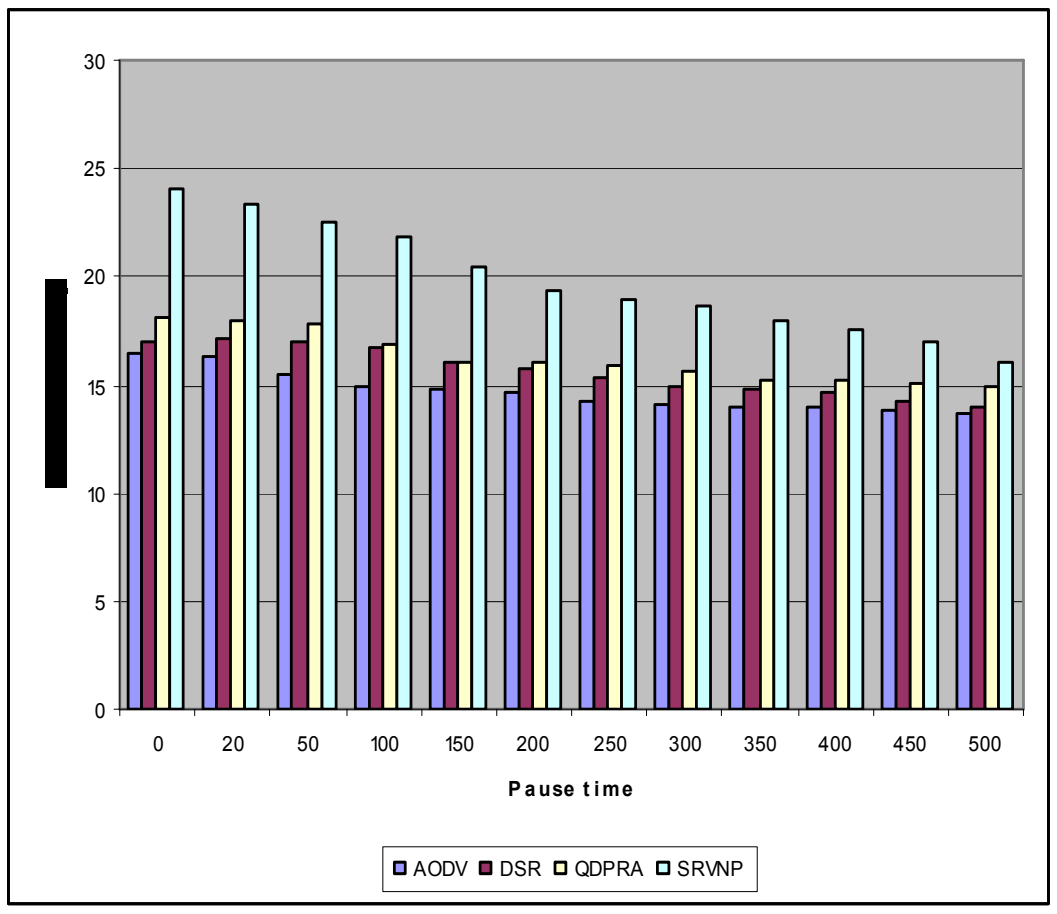

**Fig.5.** End to End Delay

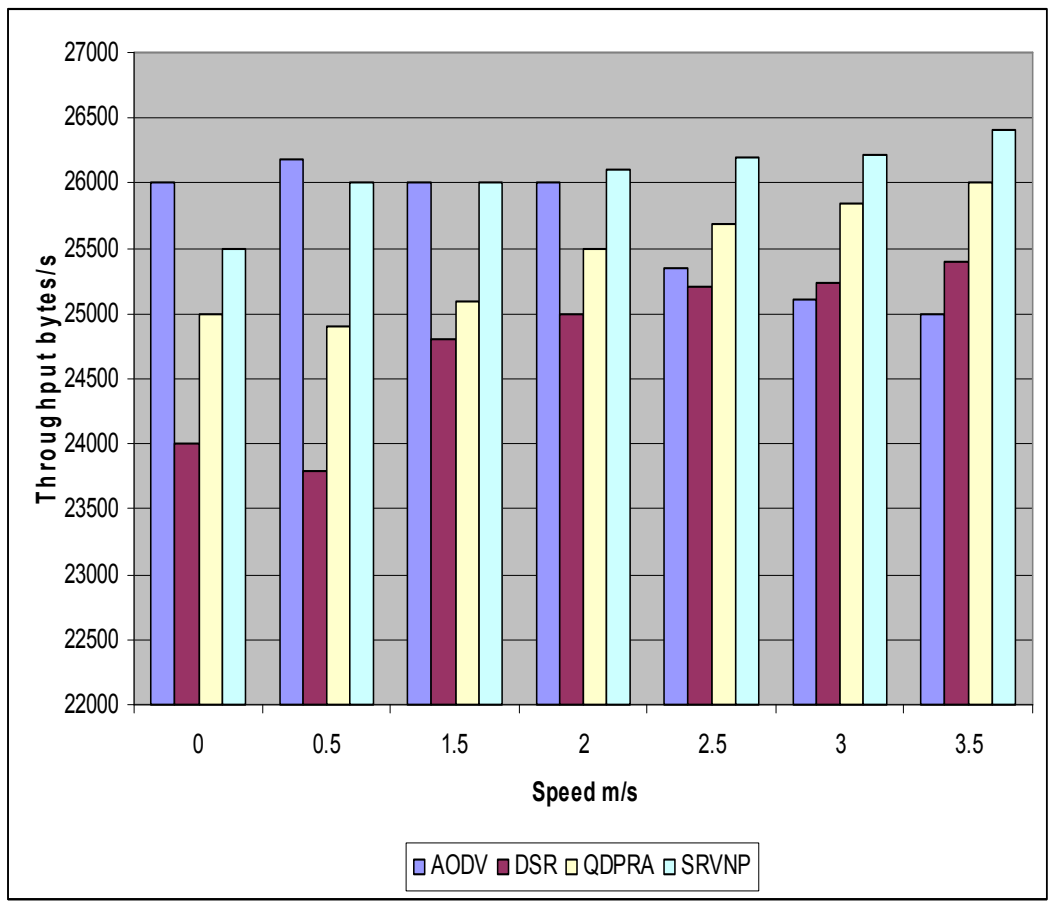

**Fig. 6.** Throughput

## 4 Conclusion

A new scheme has been presented that utilizes a mesh structure and alternate paths in case of failure. The scheme can be incorporated into any on-demand unicast routing protocol to improve reliable packet delivery in the face of node movements and route breaks. Alternate routes are utilized only when data packets cannot be delivered through the primary route. As a case study, the proposed scheme has been applied to QDPRA and it was observed that the performance improved. Simulation results indicated that the technique provides robustness to mobility and enhances protocol performance. It was found that overhead in this protocol was slightly higher than others, which is due to the reason that it requires more calculation initially for checking virtual nodes. This also caused a bit more end to end delay. The process of checking the protocol scheme is on for more sparse mediums and real life scenarios and also for other metrics like Path optimality, Link layer overhead.